\documentclass[10pt, conference, compsocconf]{IEEEtran}

\usepackage{graphicx}

\usepackage{amsmath}
\usepackage{subfigure}
\usepackage{rotating}
\usepackage{multirow}
\usepackage{array}
\usepackage{algorithm}
\usepackage{algorithmic}

\begin{document}

\title{Routing Load of Route Calculation and Route Maintenance in Wireless Proactive Routing Protocols}

\author{D. Mahmood, N. Javaid, U. Qasim$^{\ddag}$, Z. A. Khan$^{\$}$ \\
                $^{\ddag}$University of Alberta, Alberta, Canada.\\
                Department of Electrical Engineering, COMSATS\\ Institute of
                Information Technology, Islamabad, Pakistan. \\
                $^{\$}$Faculty of Engineering, Dalhousie University, Halifax, Canada.}

\maketitle
\begin{abstract}

This paper presents mathematical framework and study of proactive routing Protocols. The performance analysis of three major proactive routing protocols: Destination-Sequenced Distance Vector (DSDV) , Fish-eye State Routing (FSR) and Optimized Link State Routing (OLSR) are under consideration in this work. Taking these routing protocols into account, we enhance existing framework. In the next step we further discuss and produce analytical framework by considering variations in different network and protocol parameters. Finally, experiments are performed regarding above mentioned routing protocols followed with detailed comparison and analysis of different environments.
\end{abstract}


\maketitle

\section{Introduction}

Wireless Multi-hop Networks (WMhNs) are the solution for infrastructure less communication. They give a network of different nodes that intercommunicate with each other without any help of structured or wired devices. Hence, every node acts as a transceiver and relay a message from one node to another until message reaches its destined node. The main concept of WMhNs is to ensure freedom of communications with low costs and energy. To ensure such freedom, mobility and scalability are two major aspects which are to be tackled. No doubt such networks are gaining popularity day by day however, also give challenges for researchers in terms of efficiency.

Protocols, being vital factor that governs WMhNs communications, are functioning on different layers. Amongst all routing protocols, network layer protocols play an important role in providing smooth, and efficient functionality of a network. Actually a network layer protocol is wholly responsible for creating and maintaining all the data regarding routing of messages to their prescribed destinations.

Reactive and proactive routing protocols are two major classes of network routing protocols in WMhNs. Reactive approach is based on immediate response factor i.e., a route is searched only when it is required while proactive class is based on pre-searching of route, prior to its requirement [1]. Though, hybrid routing protocols, which actually are combination of reactive and proactive routing protocols, are also gaining popularity. Numerous experiments and simulations are undertaken with respect to proactive routing protocols while less mathematical framework is produced. In this paper we present mathematical framework that discusses behaviors of routing protocol under different environments and with variations in different network and protocol parameters. For this purpose, we take three routing protocols from proactive routing i.e., DSDV [2], OLSR [3] and FSR [4]. In proactive routing, there are two main steps involved: i.e., route calculation and route maintenance. These steps are discussed analytically and experimentally in later sections.

\section{Related Work and Motivation}
 Authors in [6] discuss and present a combined framework of reactive and proactive routing protocols. Their models deals with scalability factor. In [7], authors give analytical model which deals with effect of traffic on control overhead whereas, [8] presents a survey of control overhead of both reactive and proactive protocols. They discuss cost of energy as routing metric. Nadeem \textit{et.al.} [9], enhancing the work of [8], calculate control overhead of FSR, DSDV and OLSR separately in terms of cost of energy as well as cost of time. I.D Aron \textit{et.al} presents link repairing modeling both in local repairing and source to destination repairing along with comparison of routing protocols in [10]. X. Wu \textit{et.al}. [14] give detailed network framework where nodes are mobile and provides \emph{``statistical distribution of topology evolution''}. In [11], authors present brief understanding of scalability issues of network however, impact of topology change was not sufficiently addressed. Authors of [12] and [13] present excellent mathematical network model for proactive routing protocols. We modify the said model by adding control overhead of triggered update messages within the network.

In our work, we initially take route calculation overhead calculated by [12] as proactive control overhead. We modify given framework by adding route monitoring overhead and trigger update overhead, respectively. In next step, we calculate the aggregate routing overhead. For this purpose, different parameters of network and protocol as, number of nodes in network, route life time, periodic hello message interval for link monitoring and number of hops of network to calculate variation in routing overhead are taken. Finally, we simulate routing protocols and give a brief discussion on their respective behaviors according to different environments.

\section{Routing Protocols}
Routing protocols make the communication among different nodes possible as they are responsible of finding routes, creating routing tables and maintaining them. Besides these functionalities, they also deal with other data communication procedures. Whenever, a route is disturbed due to mobility factor or any other radio problem, routing protocol is responsible to rectify and establish the route again [1]. In this work we are confined to one prominent class of routing protocols i.e., proactive routing protocols.

\subsection{Proactive Routing}
Proactive protocols are table driven routing protocols and are meant amongst dense networks. In such protocols, routing information of next hop is preserved on the initialization of network regardless of communication requests. As, the network initializes, periodic control packets are flooded among nodes to uphold the paths or link states. In this way, they form a table on each node describing the paths to or from each node. In other words, when a network initiates, these protocols start discovering the routes within all the nodes of network regardless of use of that route.  Such procedures may cause network over burden for specific time however, reduce delay on other hand.

\subsubsection{Route Calculation}
Route calculation in proactive routing is a bit different with respect to reactive routing. In proactive routing, as the network initiates, every node hunts each and every possible destination in the network. This all information is than stored in routing tables.

\subsubsection{Route Maintenance}
As in proactive approach, every node keeps the information of all paths to every possible destination with the help of periodic messages. If a change occurs within the periodic message interval, than in some protocols, trigger message is issued. In this way, routes are maintained in proactive routing protocols.

\section{Modeling Routing Operations}
In this work, we presents a framework of proactive protocols for routing overhead. In this section, we give overall control overhead of proactive routing protocols i.e., overhead due to route calculation, un-reached destination packets, and trigger messages. Following with analyzing variations in different network and protocol parameters.  Afterwards we discuss and compare the behaviors of proactive routing protocols in different environments and scenarios .

\subsection{Proactive Route Calculation Overhead}
As, we know that in proactive approach, whenever, a network initializes, than all routes are created immediately using flooding mechanism. Routing table is updated periodically with the help of periodic messages. If any change occurs between two periodic messages, a trigger message is broadcasted, as described in DSDV [2]. In a dynamic network, there may be loss of packets due to broken links which are not updated at that vary instance. Hence, routing overhead of proactive protocols can be stated as the sum of number of packets failed to reach destination due to link breakage, periodic messages and trigger update messages. Periodic messages are issued after a specific interval of time while trigger messages are generated only when a change in topology occurs. Mathematically, we can write the above statement as:

$RO=PF+PR+TR$

$RO$ stands for routing overhead, $PF$ refers to the number of packets failed to reach destination and $PR$ stands for periodic messages while, $TR$ represents trigger messages.

Normally there are two types of errors that lead to packet failure and are discussed in detail in [9]. In either case, the probability of packet loss is increased.

\subsubsection{Route Failure Impact}

During periodic update time span ($T_{pr}$), number of packets encountering route failure is defined in [12] as:

\small
\begin{eqnarray}
RO(PF)&=& (\sum_{p_{i}\varepsilon{PA}}\sum_{r=0}^{l_{i}}Q_{r}^{l}(T_{pr})Na(T_{pr}))
\end{eqnarray}
\normalsize

\small{}
$RO(PF)$ denotes routing overhead of packet failures due to link breakage,
$Q_{r}^{l}(T_{pr})$	is probability that during first $r$ hopes, the uplink state does not change its state to down link,
$Na(T_{pr})$ specifies number of data packets arriving at time $T_{pr}$ while
$T_{pr}$ represents periodic route update time,
$L_i$ stands for length of $Pi$ ($i^{th}$ Path)and
$PA$ is set of all paths in the network.
\normalsize{}

\subsubsection{Periodic Message Overhead}

Periodic message over head in proactive routing protocols can be stated as size of routing table per periodic route update time.  While routing table size is equivalent to the size of network. Combining it with the complexity of routing overhead we get the periodic message update, as discussed in [12].

\small
\begin{eqnarray}
RO(PR)&=&\frac{kn^{3}}{BT_{pr}}
\end{eqnarray}
\normalsize

\small{}
$RO(PR)$ deonates routing overhead due to periodic updates,
$B$	is the bandwidth,
$n$	represents Number of nodes in a network and
$K$	is used to adjust routing protocol impulse factor.
\normalsize{}

\subsection{Proactive Route Maintenance overhead}
Coming to the next important aspect of aggregate routing overhead, the triggered messages, we have to understand when and how, a trigger message is updated.

Consider that there is a node in a network that moves in such a way that it changes its topology in between two periodic messages, i.e., in between $T_{pr}$ and $T_{pr}+1$, say at time $T0$. Routing protocol does not wait for next periodic message to update this change in topology instead; it immediately broadcast a triggered message. The concept of triggered message is portrayed in Fig.1.

\begin{figure}
 \begin{center}
  \includegraphics[scale=0.4]{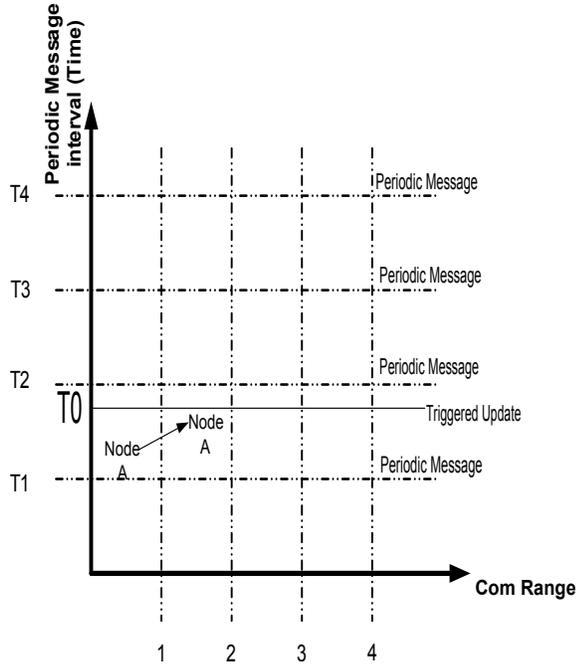}
  \caption{Node A Travels between T1 and T2 from Com Range 1 to 2, Resulting a Trigger update at T0}
 \end{center}
\end{figure}

Analytically we can express this illustration as:\\
$T_{pr} < T  < T_{pr} + 1 $.

As, discussed in [14] this notation can be expressed as:

\small
\begin{eqnarray}
RO(TR)_i&=&\frac{\left\lceil\frac{T}{T_{pr}}\right\rceil}{\frac{T}{T_{pr}}}
\end{eqnarray}
\normalsize

$RO(TR)_i$ represents routing overhead due to one triggered update message. In mathematics,  ceiling operator should be solved by taking the highest possible values.  In a network where only one node moves within a time span of $T_{pr}$ and $T_{pr}+1$, the above equation qualifies but considering a highly mobile environment where all the nodes of network are mobile, the maximum overhead due to triggered update during $T_{pr}$ and $T_{pr}+1$ is:

\small
\begin{eqnarray}
RO(TR)&=&\sum_{i=1}^{n}\frac{\left\lceil\frac{T}{T_{pr}}\right\rceil}{\frac{T}{T_{pr}}}
\end{eqnarray}
\normalsize
\small{}
$RO(TR)$ refers trigger message overhead and $T$ represents triggered update.
\normalsize{}

\subsection{Aggregate Proactive Overhead}
Combining the respective values of $Eq.1$, $Eq.2$ and $Eq. 4$  we get the analytical equation expressing the aggregate routing overhead of the network.

\tiny
\begin{eqnarray}
RO&=&(\sum_{p_{i}\varepsilon{PA}}\sum_{r=0}^{l_{i}}Q_{r}^{l}(T_{pr})Na(T_{pr}))+(\frac{kn^{3}}{BT_{pr}})+{\sum_{i=1}^{n}\frac{\left\lceil\frac{T}{T_{pr}}\right\rceil}{\frac{T}{T_{pr}}}}
\end{eqnarray}
\normalsize
Let, RO be the optimized function $``y''$ having different parameters then we get:

\tiny
\begin{eqnarray}
y(n,T_{pr},\mu_{k},T,\lambda)&=&(\sum_{p_{i}\varepsilon{PA}}\sum_{r=0}^{l_{i}}Q_{r}^{l}(T_{pr})Na(T_{pr}))(\frac{kn^{3}}{BT_{pr}})+\nonumber\\
& &{\sum_{i=1}^{n}\frac{\left\lceil\frac{T}{T_{pr}}\right\rceil}{\frac{T}{T_{pr}}}}
\end{eqnarray}
\normalsize
As, discussed in [7]:

\small
\begin{eqnarray}
Q_{r}^{l}(T_{pr})&=&1-e^{-\frac{rT_{pr}}{\mu_{k}}}
\end{eqnarray}
\normalsize
We can say that  $\lambda $ is the average number of packets arrived successfully at node, $\mu_{k}$ is the uplink time, $T$ is triggered update messages and $n$ stands for number of nodes in a network. Substituting the value from $Eq.7$ to $Eq.6$, we have:

\tiny
\begin{eqnarray}
y(n,T_{pr},\mu_{k},T,\lambda)&=&(\sum_{p_{i}\varepsilon{PA}}\sum_{r=0}^{l_{i}}1-e^{-\frac{rT_{pr}}{\mu_{k}}}\lambda(T_{pr}))+\nonumber\\
& &(\frac{kn^{3}}{BT_{pr}})+{\sum_{i=1}^{n}\frac{\left\lceil\frac{T}{T_{pr}}\right\rceil}{\frac{T}{T_{pr}}}}
\end{eqnarray}
\normalsize
To analyze variation in these parameters, we take  partial derivative of function $y$ to get:

\tiny
\begin{eqnarray}
{\partial{y}}/{\partial{T_{pr}}}&=&(C)\sum_{r=0}^{L_{avg}}{1-e^{-\frac{rT_{pr}}{\mu_{k}}}+\frac{rT_{pr}}{\mu_{k}}e^{-\frac{rT_{pr}}{\mu_{k}}}}-\nonumber\\ & & \frac{kn^{3}}{b(T_{pr})^2}+\sum_{i=1}^{n}\frac{\lceil\frac{-T}{(T_{pr})^2}\rceil+ \frac{T}{(T_{pr})^2}}{T^2/(T_{pr})^2}
\end{eqnarray}
\normalsize
where $C$ represents $PN_{avg}. \lambda$

The partial derivative with respect to the rate of packet arrival is expressed as:

\small
\begin{eqnarray}
{\partial{y}}/{\partial{\lambda}}&=&(PN_{avg})\sum_{r=0}^{L_{avg}}{T_{pr}(1-e^-\frac{rT_{pr}}{\mu_{k}})}
\end{eqnarray}
\normalsize
Similarly, if we take partial derivative with respect to $T$, we get:

\small
\begin{eqnarray}
{\partial{y}}/{\partial{T}}&=&\sum_{r=0}^{n}\frac{\lceil\frac{1}{(T_{pr})^2}\rceil- \frac{1}{(T_{pr})^2}}{T^2/(T_{pr})^2}
\end{eqnarray}
\normalsize
Rate of change in link arrival time is:

\small
\begin{eqnarray}
{\partial{y}}/{\partial{\mu_{k}}}=(PN_{avg}\lambda T_{pr})\sum_{r=0}^{L_{avg}}{-\frac{rT_{pr}}{(\mu_{k})^2}(e^\frac{-rT_{pr}}{\mu_{k}})}
\end{eqnarray}
\normalsize
It is obvious that number of nodes of a network plays a vital role to create routing overhead. We can calculate the impact of change in number of nodes of a network:

\small
\begin{eqnarray}
{\partial{y}}/{\partial{n}}&=&\frac{3kn^2}{BT_{pr}}
\end{eqnarray}
\normalsize

Considering $Eq.13$, that is partial derivative with respect to $n$, we can infer that as number of nodes of a network increases, its routing overhead increases though, if number of nodes decreases than three nodes, the overhead of the network reduces.
Assuming mobility and scalability constant,considering $Eq. 9$ and $Eq. 11$, we can say that $T_{pr}$ and $T$ are two variables that are dependent on each other. As, periodic message interval exceeds, then trigger messages are also increased. In the same way, if we reduce periodic update interval time, the ratio of generation of triggered updates is lowered if, all other parameters mainly mobility and number of nodes in a network remain constant. To further analyze this change, we take total derivative of $T_{pr}$ and $T$ variables of function $y$.

\small
\begin{eqnarray}
\frac{dy}{dT_{pr}}&=&\frac{\partial{y}}{\partial{T_{pr}}}+\frac{\partial{y}}{\partial{T}}(\frac{dT}{dT_{pr}})
\end{eqnarray}

\begin{eqnarray}
dy&=&\frac{\partial{y}}{\partial{T_{pr}}}(dT_{pr})+\frac{\partial{y}}{\partial{T}}(dT)
\end{eqnarray}
\normalsize
putting the values:

\tiny
\begin{eqnarray}
dy&=&(C)\sum_{r=0}^{L_{avg}}{1-e^{-\frac{rT_{pr}}{\mu_{k}}}+\frac{rT_{pr}}{\mu_{k}}e^{-\frac{rT_{pr}}{\mu_{k}}}}-\frac{kn^{3}}{b(T_{pr})^2}+\nonumber\\
& &\sum_{i=1}^{n}\frac{\lceil\frac{-T}{(T_{pr})^2}\rceil+\frac{T}{(T_{pr})^2}}{T^2/(T_{pr})^2}(dT_{pr})+\sum_{r=0}^{n}\frac{\lceil\frac{1}{(T_{pr})^2}\rceil- \frac{1}{(T_{pr})^2}}{T^2/(T_{pr})^2}(dT)
\end{eqnarray}
\normalsize

In static environment, longer $T_{pr}$ do not affect performance of routing protocol and favor reducing routing overhead however, if in mobile environments, longer $T_{pr}$ is used, it results in high rate of triggered messages. $Eq. 16$ shows that $T_{pr}$ and $T$ are two variables which are tied with nonlinear relationship with one another.

Considering $Eq.12$ and $Eq. 10$, it is obvious that if there is always an uplink for for the entire life of the network, or there is no periodic interval i.e., if $\mu k$ tends to infinity and $T_{pr}$ is zero, both partial derivatives with respect to $\lambda$ and $\mu k$ is zero[12]. Assuming, $\partial y /\partial T_{pr} = 0s$, we get:

\tiny
\begin{eqnarray}
C\sum_{r=0}^{L_{avg}}{(1-e^\frac{-rT_{pr}}{\mu_{k}})+e^\frac{-rT_{pr}}{\mu_{k}}}=\frac{kn^3}{(T_{pr})^2}-\frac{\lceil\frac{-T}{(T_{pr})^2}\rceil+ \frac{T}{(T_{pr})^2}}{T^2/(T_{pr})^2}
\end{eqnarray}
\normalsize

The ratio between periodic update time and uplink time can be termed as update coefficient [5]. Let us denote that update coefficient as $h= T_{pr}/ \mu{k} $ or $T_{pr}= \mu_{k}*h$. Placing the values gives us optimized network analytical model.

\tiny
\begin{eqnarray}
C\sum_{r=0}^{L_{avg}}(1-e^{-rh})+(r*h)^{-rh}=\frac{kn^3}{B(\mu_{k}*h)^2}-\nonumber\\\sum_{i=1}^{n}\frac{-\lceil\frac{T}{(\mu_{k}*h)^2}\rceil+T(\mu_{k}*h)^2}{\frac{T^2}{(\mu_{k}*h)^2}}
\end{eqnarray}
\normalsize

$Eq. 18$ shows that if average link uptime increases, update coefficient ($h$) also increases though, this increase do not linearly affect the periodic interval time. As depicted before, here again this equation shows the same, as number of nodes increases, the routing overhead is also increased non-linearly.
According to [3], there are four periodic messages in $OLSR$. $HELLO$, Topology Control messages ($TC$), Multiple Interface Declaration messages $(MID)$ and host and Network Association messages $(HNA)$. Mostly  $HELLO$ and $TC$ messages are taken into considerations. If we look into theme of $OLSR$ routing protocol, we come to know that $HELLO$ messages are used to gain the neighborhood knowledge and to select Multi-Point Relay ($MPR$) set ($MPR$ set is the only set which is allowed to retransmit or broadcast the receiving message). The nodes including in $MPR$ set are responsible for broadcasting $TC$ messages. Hence, $HELLO$ message as given in [3], is of $1$ sec while $TC$ message interval is 2 sec. In other words, [3] propose that $HELLO$ message interval should be taken as half of the $TC$ message interval.

Placing the values of $HELLO$ message and $TC$ message in composed analytical model, we get the equation:

\tiny
\begin{eqnarray}
RO_{OLSR}&=&(PN_{avg})\sum_{r=0}^{L_{i}}{(1-e^-\frac{rT_{pr}}{\mu_{k}})\lambda(T_{pr})}+\frac{kn^3}{B*H}+\nonumber\\
& &\frac{kn^3}{B*2H}+\sum_{i=1}^{n}\frac{\lceil{\frac{T}{H+2H}\rceil}}{\frac{T}{H+2H}}
\end{eqnarray}
\normalsize

$H$ =\emph{HELLO message interval},

$2H=TC$ \emph{message interval} (twice the HELLO message interval).

To analyze the rate of change in $HELLO$ and $TC$ interval, we partially derivate the above mentioned equation by $H$ and we get:

\tiny
\begin{eqnarray}
\partial{y}/\partial{H}&=&-\frac{kn^3}{B*H^2}-\frac{kn^3}{B*2H^2}+\sum_{i=1}^{n}\frac{{\lceil\frac{-T}{(H+2H^2)}\rceil}+\frac{T}{(H+2H)^2}}{\frac{T^2}{(H+2H)^2}}
\end{eqnarray}
\normalsize
With the help of same model, we can calculate desired overhead, whether to find overhead due to $HELLO$ message emission or $TC$ message overhead or the overhead due to lost packets.

\section{Simulated Results and Discussions}
Simulations of DSDV, OLSR [15] and FSR [16] are performed using NS-2. Our main concern is scalability and mobility factors in WMhNs. Simulation parameters are given below:

\emph{Simulation Parameters}\\
\small{}
1. Number of nodes  = $50$\\
2. Bandwidth        = $2\,\,Mbps$\\
3. Packet Size      = $512\,\,bytes$\\
4. Size of network  =  $1000\,\,m^2$.\\
5. Simulation setup runs on CBR\\
\normalsize{}\\
Within these parameters, we take the following three metrics.\\
\small{}
1.	Throughput\\
2.	End to end Delay\\
3.	Normalized routing Load.
\normalsize{}

\subsection{Simulation Results}
For Proactive experiments, we take FSR, DSDV and OLSR, and simulate these routing protocols with respect to mobility and scalability by taking metrics of throughput, end to end delay and normalized routing load (NRL).

\begin{figure}
\begin{center}
\includegraphics[scale=0.4]{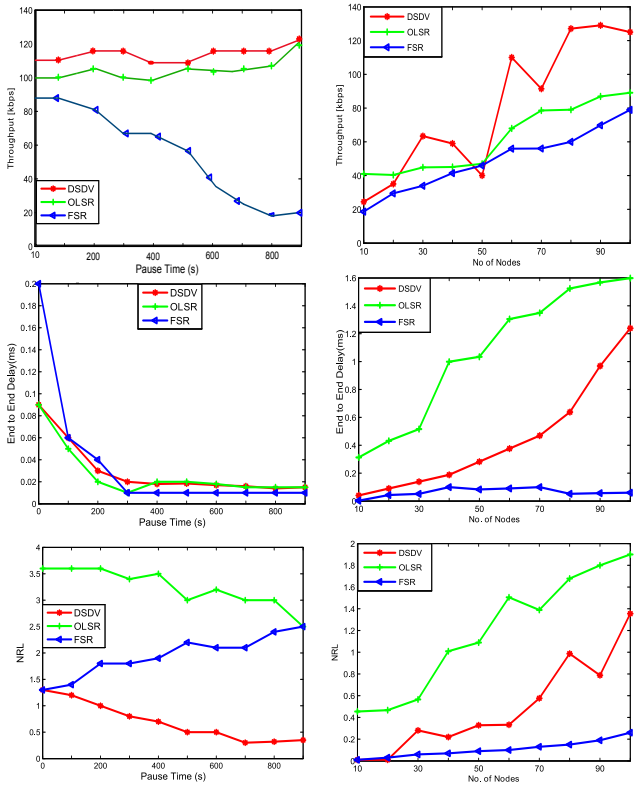}
\caption{Simulation Results of Proactive Protocols: DSDV, FSR, OLSR}
\end{center}
\end{figure}

\begin{table}
  \centering
 \caption{Comparison: Proactive Routing Protocols}
\begin{tabular}{p{1.5cm}| p{1.5cm}| p{1.5cm}| p{1.5cm} }
\hline{}
 \textbf{Feature }              & \textbf{FSR}	     & \textbf{OLSR}	       & \textbf{DSDV} \\
\hline{}
Protocol type  	& Link state	   & Link state	       & Distance Vector \\
\hline{}
Route maintained in	   &	Routing table     &	Routing table     &	Routing table \\
\hline{}
Multiple route discovery          & Yes	          & No	               & No \\
\hline{}
Multi cast	        & Yes	          & Yes               &	Yes \\
\hline{}
Periodic broad cast	    & Yes \newline(with in limited range)	& Yes	    & Yes \\
\hline{}
Topology information    & Reduced topology	& Full topology	 & Full topology \\
\hline{}
Update destination     & Neighbors	        & MPR sets	         & Source \\
\hline{}
Broadcast	           & Local/ limited	& Limited by MPR Set	& Full \\
\hline{}
Reuse of routing information		& Yes	& Yes	& yes \\
\hline{}
Route selection 	  & Shortest hop count	& Hop count	       & Shortest hop count \\
\hline{}
Route reconfiguration &	Link state mechanism  with sequence number	& Link state mechanism / Control messages send in advance	& Sequence number adopted \\
\hline{}
Route discovery packets	& Link state messages	& Via control message link sensing	& Via control messages \\
\hline{}
Limiting overhead, collision avoidance, network congestion	 &	Fisheye procedure, broadcast limited only to transmission range &	Concept of Multipoint relays	 & Concept of sequence number \\
\hline{}
Limiting overhead, collision avoidance, network congestion	& MAC layer protocols only	& MAC layer protocols only	& MAC layer protocol only \\
\hline{}
Update information	& Only neighbor information	& 2 Hop neighbor information	& By control messages \\
\hline{}


 \end{tabular}

\end{table}

\subsubsection{Throughput of Proactive Routing Protocols}

\hspace{3cm}\emph{Mobility Factor:}

DSDV outperforms among all selected protocol i.e., FSR and OLSR. Main reason of this result is basic functioning of DSDV protocol, that a packet is sent only on the best possible route due to route settling time. Moreover, un-stabilized routes that have the same sequence number in DSDV routing protocol are also advertised with delay. These features of DSDV results in accurate routing hence, throughput is increased. On the other hand, taking OLSR into account, its ability to converge declines as the mobility increases, thus results in lower throughput. Though, in static environment, due to MPR mechanism in OLSR, it gives better throughput than FSR and DSDV. Whenever, a link breaks, there is a concept of triggered messages in DSDV routing protocol that also increase the route accuracy where as in FSR there is no availability of triggered updates. OLSR triggers TC message only when status of MPRs changes.

\emph{Scalability Factor:}

In high scalabilities, OLSR outperforms among chosen protocols. OLSR uses MPR for lowering the routing overhead but periodic messages used to calculate and compute a MPR set for a node take more bandwidth. Though its throughput is more than that of DSDV however. Throughput of FSR also increases as it uses multilevel fisheye scope. This technique results in lower overhead and less consumption of bandwidth which is a major plus point for throughput. DSDV uses Network Protocol Data Units (NPDUs) for lower overhead though, triggered messages create routing overhead, consuming bandwidth and resulting in lower throughput. FSR is highly scalable as it uses different frequencies for different scopes i.e. at different time intervals.

\subsubsection{End to End Delay of Proactive Routing Protocols}
\hspace{3cm} \emph{Mobility Factor:}

DSDV proves to be the best for throughput but when considering delay, it bears the worst conditions with respect to FSR and OLSR. Moreover, delayed advertisements of unstable routes results in overall high end to end delay. In DSDV, this is done to reduce the routing overhead and provide route accuracy but it compromises on delay. In such scenario, OLSR performs better than DSDV. FSR produce the highest end to end delay among the studied protocols. As, in the basic theme of FSR, when the mobility increases, the accuracy of far away destined nodes fades. However, as the packet gets closer to destined node, the routing information gets accurate.

\emph{Scalability Factor:}

As, the network gets dense, end to end delay of discussed routing protocols i.e., FSR, DSDV and OLSR increases. FSR exchanges routing updates with its neighbors in small intervals while information shared at far away nodes has some larger interval. The network become more scalable, end to end delay increases in FSR. In DSDV, end to end delay is due to the two procedures, i.e., finding some routes and then choosing the best route. The network gets denser; end to end also increases. As in proactive nature, the information is spread in whole network. OLSR use MPRs' and in less scalable environment, end to end delay using OLSR is lowered. This is because of MPR concept that presents well organized flooding control instead of flooding a packet on whole network.

\subsubsection{Routing Load of Proactive Routing Protocols}
\hspace{3cm}\emph{Mobility Factor:}

Among the studied proactive routing protocols, OLSR generates highest routing load due to MPRs computation. DSDV again proves to be a good choice amongst FSR and OLSR in terms of routing overhead. Considering FSR, it bears lower overhead due to control and periodic messages as compared to OLSR. FSR's control messages are periodic based rather event driven based as in OLSR. This feature helps FSR to reduce routing overhead. Moreover, there is limited flooding in FSR i.e., link state information is not flooded among whole network besides, every node manage a link State table which is derived on the basis of up to date information is received. This information is not broadcasted or flooded but is shared amongst neighbors.

\emph{Scalability Factor:}

OLSR gives the highest routing overhead due to MPR computational messages and $TC$ messages. DSDV and FSR have lower overhead in dense environments. DSDV reduces overhead with the help of NPDUs. The simulated results show that FSR stands best amongst DSDV and OLSR in a dense and mobile environment in terms of overhead.

\subsection{Discsussion}
The protocol that uses minimum resources by its control packets can provide better data flow. Hence, the environments where traffic load is very high, protocols having low routing overhead will survive. If we consider scalability, in proactive routing, OLSR stands tall as it limits retransmissions due to use of MPR concept but only in dense environments. If mobility with the number of nodes of network increases, than FSR is a good choice as it generates low routing overhead that leads to high data rates within the limited bandwidth.

Considering throughput, DSDV proves itself to be the best amongst FSR and OLSR. DSDV sends a packet only on the best possible route which is verified by the protocol twice with a procedure that makes a DSDV route more accurate. This is the reason that DSDV outperforms the rest two routing protocols. OLSR's converging ability minimizes when the environment is mobile else it would prove itself to be the best due to MPR concept.

Considering routing overhead, OLSR is worst due to maximum number of periodic messages for computation of multipoint relays. DSDV proves to be a good choice considering routing overhead as well. Whereas, FSR bears lowest routing load. The feature of Fisheye scope in FSR helps in reducing the routing overhead, as, there is limited flooding i.e., link state information is not flooded among the entire network but is shared with neighbors of a scope only.

\section{Conclusion}

In this work, we give analytical model for generalized routing overhead of proactive routing protocols. For this purpose we divide the said task into small phases. In first phase we calculate the routing overhead due to route calculation and then routing overhead due to route monitoring and finally we combined them altogether that gives us the aggregate routing overhead of proactive routing protocols. Once we get the aggregate routing overhead, we then apply variations in different network and protocol's parameters and give a brief discussion on behavior of networks due to such variations. In experimental phase of our work, we simulate three most prominent proactive routing protocols keeping mobility and scalability factors into perspective along with the metrics of throughput, end to end delay and normalized routing load. We give a brief discussion of the individual behaviors of each routing protocol in different scenarios and situations.


\begin{thebibliography}{1}
\bibitem{IEEEhowto:kopka}
Ph.D. Thesis of Nadeem Javaid, ``Analysis and Design of Link Metrics for Quality Routing in Wireless Multi-hop Networks'', University of Paris-Est, 2010. http://hal.archives-ouvertes.fr/docs/00/58/77/65/PDF/TH2010PEST1028 complete
\bibitem{IEEEhowto:kopka}
C. E. Perkins and P. Bhagwat, ``Highly Dynamic Destination- Sequenced Distance-Vector Routing (DSDV) for Mobile Computers'', SIGCOMM, London, UK, August 1994, pp. 234-244.
\bibitem{IEEEhowto:kopka}
T. Clausen and P. Jacquet, ``RFC 3626, Optimized link state routing protocol OLSR'', October 2003.
\bibitem{IEEEhowto:kopka}
G. Pei, M. Gerla, and T.-W. Chen, ``Fisheye State Routing: A Routing Scheme for Ad Hoc Wireless Networks'', Proceedings of ICC 2000, New Orleans, LA, Jun. 2000.
\bibitem{IEEEhowto:kopka}
Mohammad Naserian, Kemal E. Tepe, Mohammed Tarique, ``Routing Overhead Analysis for Reactive Routing Protocols in Wireless Ad Hoc Networks'', Wireless And Mobile Computing, Networking And Communications, 2005. (WiMob'2005), IEEE International Conference on, vol.3, no., pp. 87- 92 Vol. 3, 22-24 Aug. 2005.
\bibitem{IEEEhowto:kopka}
Hui Xu, et.al, `` A Unified Analysis of Routing Protocols in MANETs'', IEEE Transactions on Communications, VOL. 58, NO. 3, March 2010.
\bibitem{IEEEhowto:kopka}
Nianjun Zhou, et.al., ``The Impact of Traffic Patterns on the Overhead of Reactive Routing Protocols'', IEEE Journal on Selected Areas in Communications, VOL. 23, NO. 3, March 2005
\bibitem{IEEEhowto:kopka}
Jacquet, P. and Viennot, L., ``Overhead in mobile ad-hoc network protocols'', RAPPORT DE RECHERCHE-INSTITUT NATIONAL DE RECHERCHE EN INFORMATIQUE ET EN AUTOMATIQUE, 2000.
\bibitem{IEEEhowto:kopka}
Nadeem Javaid et al. ``Modeling routing overhead Generated by Wireless Proactive Routing Protocols'', USA Globecom 2011.
\bibitem{IEEEhowto:kopka}
I.D. Aron and S. Gupta, ``Analytical Comparison of Local and End-to-End Error Recovery in Reactive Routing Protocols for Mobile Ad Hoc Networks'', Proc. ACM Workshop Modeling, Analysis, and Simulation of Wireless and Mobile Systems (MSWiM), 2000.
\bibitem{IEEEhowto:kopka}
N. Zhou and A.A. Abouzeid, ``Routing in ad hoc networks: A theoretical framework with practical implications'',INFOCOM 2005, pages 1240-1251, Miami, 2005.
\bibitem{IEEEhowto:kopka}
Pan-long Yang, Chang Tian and Yong Yu, ``Analysis on optimizing model for proactive ad hoc routing protocol'', Military Communications conference, 2005. MILCOM 2005. IEEE , vol., no., pp.2960-2966 Vol. 5, 17-20 Oct. 2005.
\bibitem{IEEEhowto:kopka}
Ben Liang and Zygmunt J. Hass, ``Optimizing Route-Cache Lifetime in Ad Hoc Networks'', Proceedings of the 22nd IEEE INFOCOM April 1st 3rd ,2003.
\bibitem{IEEEhowto:kopka}
Xianren Wu, Sadjadpour, H.R., Garcia-Luna-Aceves, J.J., ``Routing Overhead as A Function of Node Mobility: Modeling Framework and Implications on Proactive Routing'', Mobile Adhoc and Sensor Systems, 2007. MASS 2007. IEEE Internatonal Conference on , vol., no., pp.1-9, 8-11 Oct. 2007.
\bibitem{IEEEhowto:kopka}
University of Murcia,``MASIMUM (MANET Simulation and Implementation at the University of Murcia)'', 2009.
\bibitem{IEEEhowto:kopka}
The Institute of Operating Systems and Computer Networks,``FSR''.

\end{thebibliography}
\end{document}